\documentclass[a4paper,twocolumn]{esapub} 
\usepackage{natbib}

\title{INTEGRAL observations of Galaxy Clusters}
\author[1]{P. Goldoni}
\author[1]{A. Goldwurm}
\author[1]{P. Laurent}
\author[2]{M. Cass\'e}
\affil{DAPNIA/Service d'Astrophysique, CEA/Saclay,
F-91191 Gif-Sur-Yvette, France}
\affil{IAP/Paris}
\author[1]{J. Paul}
\author[3]{C. Sarazin}
\affil{University of Virginia}

\input{psfig.sty}
\begin{document}


\maketitle

\begin{abstract}
Cluster of galaxies are the largest concentrations of visible mass in
the Universe and therefore a fundamental topic of cosmology and
astrophysics. Recent radio, EUV, and X-ray observations suggest that
clusters contain large populations of diffuse nonthermal relativistic
and/or superthermal particles.  These particles may be produced by
acceleration in cluster merger shocks, AGNs, and/or supernovae in
cluster galaxies.  Models for the nonthermal populations in clusters
indicate that they should produce substantial hard X-ray and $\gamma$
luminosities. The possible role of nonthermal particles in the
dynamics of clusters is one of the greatest uncertainties in their use
as cosmological probes.

INTEGRAL offers, for the first time, the possibility of simultaneous
medium resolution imaging ($\sim$ 12 arcmin) and high resolution spectroscopy
($\Delta$ E/E $\sim$ 2 keV @ 1.3 MeV) with exceptional sensitivity in the hard
X-ray/soft $\gamma$-ray band.  The spatial resolution will allow discrete
sources, such as AGNs, to be separated from the diffuse emission of the
cluster.  For nearby clusters, the spatial distribution of emission can
be determined and compared to models for the nonthermal particle
populations and observations in other wavebands.  The spectral
capabilities may allow different components of the nonthermal
population to be detected separately.  We present simulations of
INTEGRAL observations of nearby galaxy clusters that show its
capability for detecting the different phenomena responsible of clusters
emission.

\end{abstract}

\section{Introduction}
\noindent Clusters of Galaxies have been thoroughly investigated along
the electromagnetic spectrum. Their properties give us valuable hints on
the evolution of the Universe and they are one of the
best cosmological probes in our possession. Besides the thermal gas
which dominates soft X-ray observations, it has long been believed
that clusters of galaxies harbor a large populations of non thermal
particles. Clusters should be effective traps for cosmic ray particles
and the high temperature gas in the IntraCluster Medium (ICM) indicates
that strong shock are operating inside it. If diffuse gas undergoes
a strong shock, a portion of the shock energy goes into accelerating
relativistic particles.

\noindent Relativistic electrons emit radiation either directly or
by Inverse Compton interaction on the ambient photons. On the other hand
relativistic protons collide on the nuclei in the intracluster
medium producing radioactive isotopes and nuclear lines. The detection
of diffuse {\sl radio} halos in the cluster cores (see e.g Giovannini 
et al. 1993) confirms this scenario.

\noindent Recently this topic has been revived by the detection of
EUV and hard X-ray excesses (Bowyer et al. 1999, Sarazin \& Lieu 1998,
Fusco-Femiano et al. 1999, Rephaeli et al. 1999, Kaastra et al. 1998,
Fusco-Femiano et al. 2000) in three nearby clusters: Coma, A2199 and A2256.
The origin of this radiation has been widely discussed but no certain origin
can be up to now determined (Sarazin 1999). The debate on the origin of hard 
X-ray excesses has been up to now hampered by the lack of imaging
instruments in this energy domain.

\noindent Indeed at the flux level reported in the literature ($\sim
$ few $\times$ 10$^{-11}$ erg cm$^{-2}$ s$^{-1}$ in the 20-100 keV
energy band, Bassani et al. 1999), there is a non negligible ($\sim$ 10
$\%$) possibility of source confusion (mainly with AGNs) in the
relatively large ($\sim$ 1 degree) field of view of actual instruments.
Also the parameters and even the detection of the hard X-ray excess are
highly dependent on the parameters of the spectral fit performed on the
cluster soft X-ray emission.

\noindent The INTEGRAL satellite, the next ESA hard X-ray/soft $\gamma$-ray
mission, will be able to advance our knowledge in these topics thanks to its
exceptional sensitivity and imaging and spectral capabilities, unprecedented
in this energy domain. The imaging capability of the IBIS imager will allow
to identify the site of the emission and therefore the distribution
of high energy electrons in the cluster. On the other hand the SPI spectrometer
will allow to probe the acceleration of nuclei and their subsequent interaction 
with the ambient medium. We will show in the following the results of 
simulations of INTEGRAL observations which allow us to evaluate its impact
on this problem.

\section{The INTEGRAL satellite and its instruments}

\noindent INTEGRAL is a 15 keV-10 MeV $\gamma$-ray mission with concurrent
source monitoring at X-rays (3-35 keV) and in the optical range (V, 500-
600 nm). All instruments are coaligned and have a large FOV, covering
simultaneously a very broad range of sources. The INTEGRAL payload consists
of two main $\gamma$-ray instruments, the spectrometer SPI and the imager IBIS,
and of two monitor instruments, the X-ray monitor JEM-X and the Optical
Monitoring Camera OMC.


\noindent The Imager on Board Integral Satellite (IBIS) provides diagnostic capabilities
of fine imaging (12' FWHM), source identification and spectral sensitivity
to both continuum and broad lines over a broad (15~keV--10~MeV) energy range.
It has a continuum sensitivity of 2~10$^{-7}$~ph~cm$^{-2}$~s$^{-1}$ at 1~MeV
for a 10$^6$ seconds observation and a spectral resolution better than 7~$\%$
@ 100~keV and of 6~$\%$ @ 1~MeV. The imaging capabilities of the IBIS are 
characterized by the coupling of its source discrimination capability
(angular resolution 12' FWHM) with a field of view (FOV) of 9$^\circ$
$ \times $ 9$^\circ$ fully coded, 29$^\circ$ $ \times $ 29$^\circ$ partially coded FOV.


\noindent The spectrometer SPI will perform spectral analysis of $\gamma$
ray point sources and extended regions with an unprecedented energy
resolution of $\sim$ 2 keV (FWHM) at 1.3 MeV. Its large field of view
(16$^{\circ}$ circular) and limited angular resolution ( 2$^{\circ}$ FWHM)
are best suited for diffuse sources imaging but it retains nonetheless the
capability of imaging point sources. It has a continuum sensitivity of
7 $\times$ 10$^{-8}$ ph cm$^{-2}$ s$^{-1}$ at 1 MeV and a line sensitivity
of 5$\times$ 10$^{-6}$ ph cm$^{-2}$ s$^{-1}$ at 1 MeV, both 3$\sigma$ for
a 10$^6$ seconds observation.


\noindent The Joint European Monitor JEM-X supplements the main INTEGRAL instruments
and provides images with 3' angular resolution in a 4.8$^{\circ}$ fully coded
FOV in the 3-35 keV energy band. The Optical Monitoring Camera (OMC) will
observe the prime targets of INTEGRAL main $\gamma$ ray instruments. Its
limiting magnitude is M$_V$ $\sim$ 19.7 (3$\sigma$, 10$^3$ s).
The wide band observing opportunity offered by INTEGRAL provide for the first 
time the opportunity of simultaneous observing over 7 orders of magnitude.

All INTEGRAL instruments will be important in defining the emission from
galaxy clusters, however in this paper we will focus on IBIS contribution
in defining the cluster shape at $>$ 20 keV energies and in identifying
the origin of non thermal emission. We will show that IBIS will be able
to determine the position of the $>$ 20 keV emission inside the cluster
and therfore to give precious hints on the acceleration processes in the
ICM.

\section{Simulations}


\noindent We simulated an IBIS 10$^6$ seconds observation of the Coma
cluster, the pointings were centered on the cluster with a standard
3 x 3 rectangular dithering pattern. SAX/PDS and RXTE observations of
this cluster resulted in a detection of a 20-100 keV flux of about 2.2
mCrab in their 1.3$^{\circ}$ and 1$^{\circ}$ FWHM degree fields of view 
(Fusco-Femiano et al. 1999, Rephaeli et al., 1999). The origin of this
emission is however unclear, as this kind of instrument is only capable of recording the {\sl total} emission in the region without discriminating
between the components. The spectral fit performed on the data required
a non thermal component in excess of the thermal one. In Figure 1 we show
a simulated spectrum of the 0.1-100 keV emission detected in Coma Cluster
region by previous observations (Hughes et al. 1993, Rephaeli et al.
1999). The spectrum is the composition of a Raymond-Smith thermal model
with kT = 9 keV and a non thermal power law with photon index $\alpha$ =
2.35.

\noindent Thermal emission clearly dominates the total flux at energies
smaller than 50 keV. At higher energies however the emission is fainter
and the situation is much less clear. From the plot it is evident that
the parameters of the non thermal power law can be strongly influenced
by a change in the parameters of the much stronger thermal component.
Also a serendipitous AGN in the instrument FOV could substantially
contribute to the total flux. The value of this component is difficult
to assess as the spectral and temporal properties of AGNs are relatively
unknown in this energy range. The strongest AGN in the Coma Cluster region
is the Seyfert I galaxy X Comae (Bond \& Sargent 1973), located $\sim$ 30
arcmin from the cluster centre. Extrapolating the ROSAT detection by
Dow \& White (1995) we have estimated a 20-60 keV X-ray flux of $\sim$
0.2 mCrab.

\begin{figure}
\centerline{\psfig{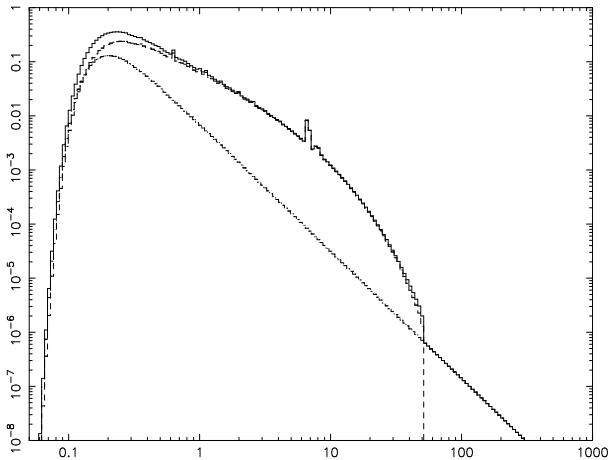}}

\caption{Simulated 0.1-100 keV spectrum of the Coma cluster region
obtained using data from non imaging observations of Hughes et al. 1993
(GINGA) and Rephaeli et al. 1999 (RXTE). Two components are visible: the
thermal Raymond-Smith model dominates at lower energy up to $\sim$ 50
keV while the non thermal power law only is present at higher energies}

\end{figure}

\noindent 
We choose to simulate IBIS images of Coma cluster and X Comae in two
energy bands: 20-60 keV ('thermal') and 40-120 keV ('non thermal').
Using the spectrum shown in Figure 1, the cluster emission in the low
energy band is $\sim$ 1.6 mCrab, while in the high energy band is
$\sim$ 0.4 mCrab. At low energies, we simulated its shape as a symmetric
gaussian with 30' and 60' diameter FWHM, similar to what found in ROSAT
images (see e.g. White et al. 1993). Our simulation assumes an isothermal
cluster for simplicity, see Honda et al. 1996 for an analysis of Coma
temperature structure. The non thermal emission structure is not determined,
it should be linked to clusters' radio halo whose radius varies with
frequency. We consider that its dimensions should be somewhere in between
the 30-40 arcmin radius at 150 MHz (Cordey 1985) and the $\sim$ 10 arcmin
radius at 1.4 GHz (Kim et al. 1990). We simulated two nonthermal emission
configurations: one gaussian with respectively 30' and 15' diameter FWHM.
A smaller diameter being not resolvable by IBIS, it would result in a
point source for the instrument. In the following we report the results
of our simulations, we remark that the confidence level we report
for our detections are preliminary.

\noindent Figure 2 and 3 show the result of our simulations in the 20-60
keV energy band, the cluster is clearly detected at a significance level
of 20 and 60 $\sigma$ respectively while X Comae is detected in both cases
at $\sim$ 15 $\sigma$ level. The cluster's {\sl extended} nature is clearly
visible in our images thus demonstrating IBIS' capabilities. The cluster
appear in a much clearer way in Figure 3 due to the much higher surface
brightness. A similar image will be the first at such energies and it will
help determine the temperature structure of the cluster giving basic
information on the distribution of $>$20 keV gas in the cluster. Moreover
it should be noted that the spectral analysis will not be hampered by
the presence of a contaminating source which would be readily identified.

\begin{figure}

\centerline{\psfig{figure=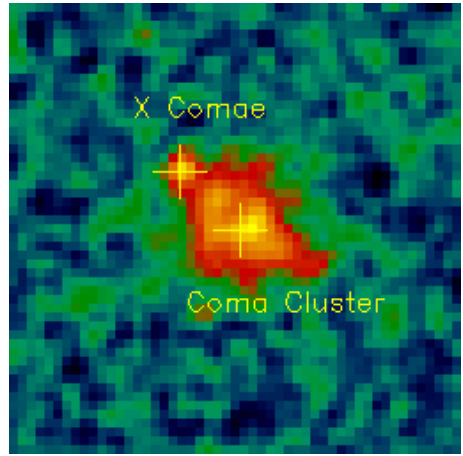,height=60mm,width=60mm}}
\caption{20-60 keV color-coded image of a simulated 10$^6$ sec observation
of the Coma cluster. X Comae flux is 0.2 mCrab while the cluster flux is
2.0 mCrab, its shape is a gaussian shape of 60' FWHM. The cluster is clearly
visible as an extended source, and its shape its clearly distorted by
the presence of X Comae on the upper left.}

\end{figure}

\begin{figure}

\centerline{\psfig{figure=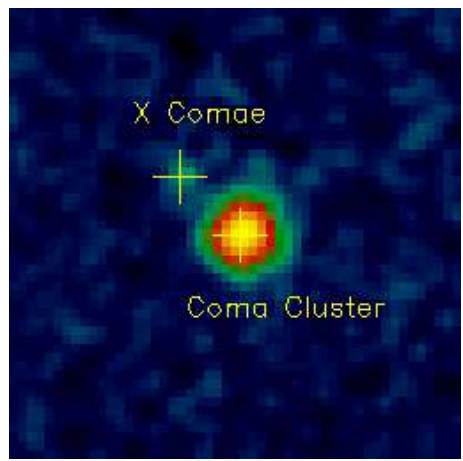,height=60mm,width=60mm}}
\caption{20-60 keV color-coded image of a simulated 10$^6$ sec observation
of the Coma cluster. Fluxes of sources are the same as above, the
cluster is a gaussian of 30' FWHM. The cluster is again clearly
visible as an extended source, while X Comae is a fainter source on
the side which contributes a sizable flux.}

\end{figure}

\noindent Figure 4 and Figure 5 show the result of our simulation in
the 40-120 keV energy band in the two cases defined above. In both
cases the cluster is detected but at a 5 $\sigma$ level in the first
case and at a 10 $\sigma$ level in the second. Inspection of the image
clearly shows that the cluster detection is much more difficult as 
surface brightness is much smaller in the second case. Background
fluctuations could make the detection more difficult in this case, 
obliging to employ more sophisticated detection methods. As expected
X Comae is proportionally stronger at a $\sim$ 4 $\sigma$ level.

\begin{figure}

\centerline{\psfig{figure=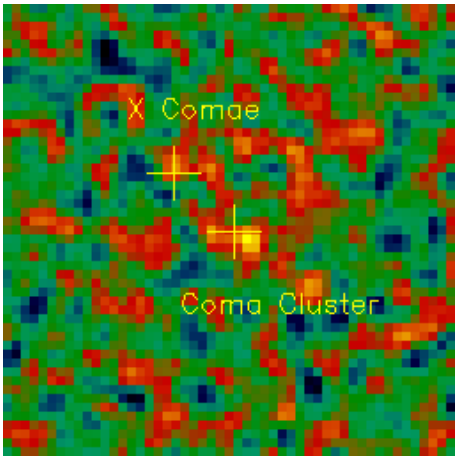,height=60mm,width=60mm}}
\caption{40-120 keV color-coded image of a simulated 10$^6$ sec 
IBIS observation of the Coma cluster. The cluster has been simulated
with a gaussian with 30' FWHM. The cluster is fainter and it is
detected at a 5 $\sigma$ level. X Comae is proportionally
stronger at a 4 $\sigma$ level.}

\end{figure}

\begin{figure}

\centerline{\psfig{figure=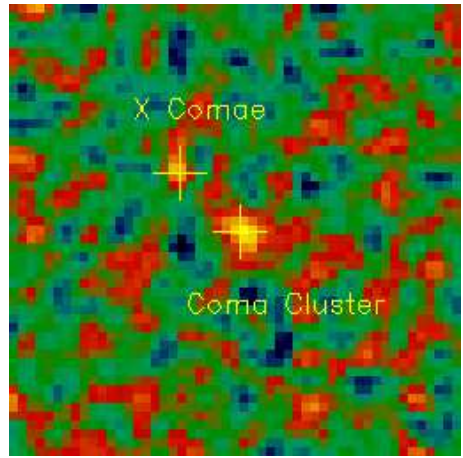,height=60mm,width=60mm}}
\caption{40-120 keV color-coded image of a simulated 10$^6$ sec 
IBIS observation of the Coma cluster. The cluster has been simulated
with a gaussian with 15' FWHM and is much more clearly visible
than in Figure 4 as the flux is spreaded on a smaller surface
The detection significance is around $\sim$ 10 $\sigma$.
X Comae stays at 4 $\sigma$ level.}

\end{figure}

\section{Conclusions}

\noindent We simulated a 10$^6$ seconds observation of the Coma cluster
with the IBIS instrument onboard the INTEGRAL satellite. We simulated
a configuration comprising an AGN (X comae) in the Fully Coded Field of View
of the instrument and we took the cluster emission parameter from Hughes
et al. (1993) and Rephaeli et al. (1999). We demonstrated that IBIS
is fully able to detect and separate the two components of the emission.

\noindent In the two images in the 20-60 keV energy band, the cluster
is clearly detected as an {\sl extended} source at a 20-60$\sigma$ level 
depending on the assumed shape. This will be the first image of the
cluster at these energies and in itself will help better define the
temperature distribution. Together with JEM-X data this should
give a complete picture of the clusters' thermal emission.

\noindent At higher energies, in the 40-120 keV energy band, the
results strongly depend from the unknown distribution of the clusters'
non thermal emission. If its dimension is $<$ 15 arcminutes
FWHM (600 kpc), then it can be detected at $\sim$ 10 $\sigma$ level.
However if the emission is truly "diffuse" greater than IBIS
resolving power, then the detection is much more difficult.



\end{document}